\documentstyle[epsfig,aps,preprint]{revtex}
\tighten

\def\la{\mathrel{\mathchoice
{\vcenter{\offinterlineskip\halign{\hfil
$\displaystyle##$\hfil\cr<\cr\sim\cr}}}
{\vcenter{\offinterlineskip\halign{\hfil$\textstyle##$\hfil\cr<\cr\sim\cr}}}
{\vcenter{\offinterlineskip\halign{\hfil$\scriptstyle##$\hfil\cr<\cr\sim\cr}}}
{\vcenter{\offinterlineskip\halign{\hfil$\scriptscriptstyle##$\hfil\cr<\cr\sim
\cr}}}}}

\def\ga{\mathrel{\mathchoice
{\vcenter{\offinterlineskip\halign{\hfil
$\displaystyle##$\hfil\cr>\cr\sim\cr}}}
{\vcenter{\offinterlineskip\halign{\hfil$\textstyle##$\hfil\cr>\cr\sim\cr}}}
{\vcenter{\offinterlineskip\halign{\hfil$\scriptstyle##$\hfil\cr>\cr\sim\cr}}}
{\vcenter{\offinterlineskip\halign{\hfil$\scriptscriptstyle##$\hfil\cr>\cr\sim
\cr}}}}}

\begin{document}

\title{Chaos in a well : Effects of competing length scales}

\author {R. Sankaranarayanan\footnote{sankar@prl.ernet.in}, 
A. Lakshminarayan\footnote{arul@prl.ernet.in} and 
V. B. Sheorey\footnote{sheorey@prl.ernet.in}}
\address{Physical Research Laboratory,\\
Navrangpura, Ahmedabad 380 009, India.}
\maketitle

\begin{abstract}
A discontinuous generalization of the standard map, which  
arises naturally as the dynamics of a periodically kicked
particle in a one dimensional infinite square well
potential, is examined.
Existence of competing length scales, namely the width of
the well and the wavelength of the external field, introduce novel
dynamical behaviour. Deterministic chaos induced 
diffusion is observed for weak field strengths
as the length scales do not match. This is related to an
abrupt breakdown of rotationally invariant curves and in particular KAM
tori.  An approximate stability theory is derived wherein the usual
standard map is a point of ``bifurcation''.\\[10pt]
PACS number(s) : 05.45Ac, 45.05.+x, 47.52.+j \\
Keywords : Chaos
\end{abstract}

\section {Introduction}

The construction and study of area preserving mappings has led to a
deeper understanding of apparently complex dynamics, especially of
Hamiltonian chaos. The maps range from abstract models \cite{arnold}
such as the
cat maps and the baker map to more generic ones.  One of the most well
studied of such generic mappings is the standard map \cite{chirikov},
which has also been investigated extensively in its quantum version
\cite{izrailev}. The classical map on the cylinder displays a range of
dynamical behaviours, and when completely chaotic, diffusive random
walks in momentum take place.  Recent experiments using trapped
ultra-cold sodium atoms in pulsed laser fields have probed this model,
as a kicked rotator, and verified the central phenomenon of quantum
localization of momentum diffusion \cite{moore}.

In this Letter we study the classical dynamics of a simple
generalization of the standard map that naturally arises from the
dynamics of a particle trapped in a one dimensional infinitely deep
well. The virtue of this generalization is that on the introduction of
a competing length scale novel dynamical behaviour manifests, leading
to large scale diffusive processes even for small external field
strengths. This places the standard map within a family of generally
discontinuous area preserving mappings; the points where there is
continuity corresponding to the usual standard map. 

We study the novel dynamical changes by simple methods exploiting
separation of time scales in non-autonomous equations. The failure of
the Poincar\'{e}-Birkhoff theorem leads to the formation of entirely
stable islands and cantori or entirely unstable and chaotic orbits. It
is also to be pointed out that this can be one of the methods of
controlling or enhancing chaos.

The classical motion of a particle in an infinite square well
potential in the presence of a uniform monochromatic external field
has been studied previously \cite{lin}. In this case diffusion was
observed over a limited range of energies and there is no issue of
competing length scales. Also, no analytic form of the stroboscopic
mapping is derivable. We have considered below a space variation of
the external field that is also pulsed in time, the pulsing leading to
analytically derivable mappings. More recently a special case of this
system was studied \cite{bambi} and quantized, 
where (in the chaotic regime) an 
important effect was found, namely delocalization of eigenstates as opposed
to the well known exponentially localized states of the kicked rotator.  Our
independent and parallel studies are a generalization and in this
Letter we present the classical aspects as we feel that these
deserve a more complete understanding. 

Recent advances in semiconductor physics, makes it possible to construct 
wells on atomic scales. Also, developments in investigating 
the quantum nature of electrons in a finite quantum well have been achieved, 
using tunnel-current spectroscopy, when the corresponding classical system 
shows chaotic behaviour \cite{wilk} (see references therein). We hope that 
the study of novel dynamical behaviours in simple models, as the 
generalization of the standard map, will reflect on some aspects of such 
experimentally realizable systems.

\section {Kicked particle in a well}

We consider a particle of mass $m$, trapped in an one dimensional infinite 
square well potential $V_{sq}(x)$ of width $2a$. There is 
an external field $V(x)$ that is  periodically  pulsed with period $T$.
The Hamiltonian is  

\begin{equation}
H = {p^2 \over 2m} + V_{sq}(x) + V(x)\sum_{j=-\infty}^{\infty} 
\delta\left(j-{t \over T}\right) 
\label{h}
\end{equation}
\noindent where $x,p$ are the position and momentum of the particle
respectively.  We consider below $V(x)= \epsilon \cos(2\pi
x/\lambda)$. Here $\epsilon$ and $\lambda$ are the amplitude and
wavelength of the external field respectively. The effect of the
pulsed field in $H$ is the same as that of an infinite number of
travelling waves with identical amplitudes and frequencies which are 
multiples of $2\pi/T$ (the pulse frequency). 

The stroboscopic map, relating the dynamical variables immediately after 
successive kicks, can be derived in a standard manner and
results in the following dimensionless form:
\begin{equation}
\begin{array}{lll}
X_{n+1}&=&{(-1)}^{M_n} (X_n + P_n) + {(-1)}^{M_n+1} \mbox{Sgn}(P_n) M_n \\[6pt]
P_{n+1}&=&{(-1)}^{M_n} P_n + (K/2 \pi) \sin(2 \pi RX_{n+1}).
\end{array}
\label{map}
\end{equation}
\noindent where $M_n=\left[\mbox{Sgn}(P_n)(X_n+P_n)+{1 \over 2}\right]$, 
is the number of bounces of the particle at the walls during the  
interval between the $n$th kick and the $(n+1)$th kick. 
Here Sgn(..) and $[..]$ stand for sign and integer 
part of the argument respectively.
The dimensionless quantities are defined as
\begin{equation}
X_n = {x_n \over 2a}, \;\;\;\;\; 
P_n = {p_nT \over 2am}, \;\;\;\;\;
K = {2 \epsilon {\pi}^2 T^2 \over am \lambda}, \;\;\;\;\; 
R = {2a \over \lambda}.
\end{equation}
while the dynamical variables $x_n$, $p_n$ are the position and
momentum of the particle just after the $n$th
kick. Although Sgn($P_n$) is discontinuous and undefined for $P_n=0$,
one can see easily from the map that either of the values ($\pm 1$)
can be taken for $\mbox{Sgn}(P_n)$ as this does not alter the
dynamics. We refer to the mapping (\ref{map}) as the {\it well map}. 
Note that $K$ and $R$ are 
the only two effective parameters of the well map. Here $K$ is related to 
the strength of the external field; $R$ is the ratio of the width of the well 
to the wavelength of the external field. We do not impose any constraints 
on $R$, for example those while considering standing waves in a cavity, 
but allow it all values. Evidently,  $|X_n| \le 1/2$ or $|x_n| \le a$ as
the particle motion is confined between two rigid walls.

We now relate the well map to a generalized standard map.
The motion of a free particle inside the potential well $V_{sq}(x)$
is related to the motion of a free rotator, the difference being one of
boundary conditions. Both free motions are same unless the particle 
hits the walls. The correspondence between them is made explicit in the
following way. Let us consider the dynamics of a free rotator in discrete
unit time steps, which is the map relating successive angles ${\theta}_n$ and 
angular momenta $J_n$.
\begin{equation}
{\theta}_{n+1}={\theta}_n+J_n \hspace*{0.2in} 
\hbox{(mod 1)} \;\;\;\;\;;\;\;\;\;\; J_{n+1}=J_n.
\label{rotor}
\end{equation}
\noindent Here the motion is confined to a cylinder $[-{1/2},{1/2}) \times 
(-\infty,\infty)$. 
Similarly the discretized free motion in a well of unit width is 
\begin{equation}
X_{n+1}={(-1)}^{M_n} (X_n + P_n) + {(-1)}^{M_n+1} 
\mbox{Sgn}(P_n) M_n \;\;\;\;\; ; \;\;\;\;\;
P_{n+1}={(-1)}^{M_n} P_n
\label{particle}
\end{equation} 
Let us denote $S_n \equiv({\theta}_n,J_n)$,  and $W_n\equiv(X_n,P_n)$ .
If we have the same initial condition for both the maps 
(\ref{rotor}) and (\ref{particle}) i.e., $W_0=S_0$, we have the following 
relation after the $n$th iterate:
\begin{equation}
W_n\, =\, (-1)^M \, S_n
\label{state}
\end{equation} 
\noindent where $M={\sum}_{i=0}^{n-1}M_i$, is the {\it total} number of
bounces of the particle after $(n-1)$ iterates. 
In the rest of the paper we consider the mapping
\begin{equation}
\begin{array}{lll}
J_{n+1}&=&J_n+(K/2\pi)\sin(2\pi R{\theta}_n) \\[6pt]
{\theta}_{n+1}&=&{\theta}_{n}+J_{n+1} \hspace*{0.6cm} (\mbox{mod}\;\;1).
\end{array}
\label{gsm}
\end{equation}
It is easy to see that we have the same relation
(\ref{state}) between the well map (\ref{map}) and the map defined 
in (\ref{gsm}) (more precisely the time reversal of this map). 

We refer to the map defined in (\ref{gsm}) as the {\it Generalized
Standard Map} (GSM). From the above description it is clear that
quantitative dynamical features are same for the well map and the
GSM. In attempting to analyze the dynamics of the well map, it is
sufficient to understand the dynamics of the GSM. Also the map
(\ref{gsm}) is easier to handle than the well map, as it is only a
slightly generalized version of the well studied standard map ($R=1$)
\cite{chirikov}.  When it is stated that the standard map is a one
parameter system the implicit assumption is that there is only one
length scale. Here there are naturally {\em two} length scales whose
ratio is the cause of many interesting effects. Both the maps
(\ref{map}) and (\ref{gsm}) have reflection symmetry about their
origin. Below we discuss various dynamical behaviour of the GSM in
more detail. Throughout this paper we consider $R$ to assume positive
real values only. 

\section {Dynamical features}

For the standard map (GSM with $R=1$), when $K \ll 1$, the dynamics is
quasi-regular and the phase space is filled with a large number of KAM
tori which are {\it rotationally invariant circles} \cite{meiss}
extending across the phase plane.  These
closed loops ${\cal C}$ encircle the cylinder and are invariant,
meaning that ${\cal T}{\cal C}={\cal C}$, where ${\cal T}$ is an area
preserving {\it continuous} transformation on the cylinder (for
instance Eqn.(\ref{gsm}) with $R=1$). KAM tori are the principal
barriers for the unstable orbits to diffuse in the phase space. As $K$
increases there is a {\it smooth} transition from regular to chaotic
behaviour with a reduction in the number of KAM tori. Dynamical
changes of KAM tori with increase of $K$ are discussed in
\cite{shenker}. At $K \simeq 1$ all the KAM tori disappear from the
phase space \cite{greene}; the dynamics is chaotic and diffusive for
$K \gg 1$.  Note that the GSM is periodic in $J$ and $\theta$ with
unit period. For integer values of $R$, it is
continuous and takes the form of standard map with $K \rightarrow KR$.
In this case dynamical transitions are identical with that of the
standard map. As the effective perturbation here is $KR$, higher
integer values of $R$ also make the dynamics more chaotic.

\begin{figure}[ht]
\centerline{\psfig{figure=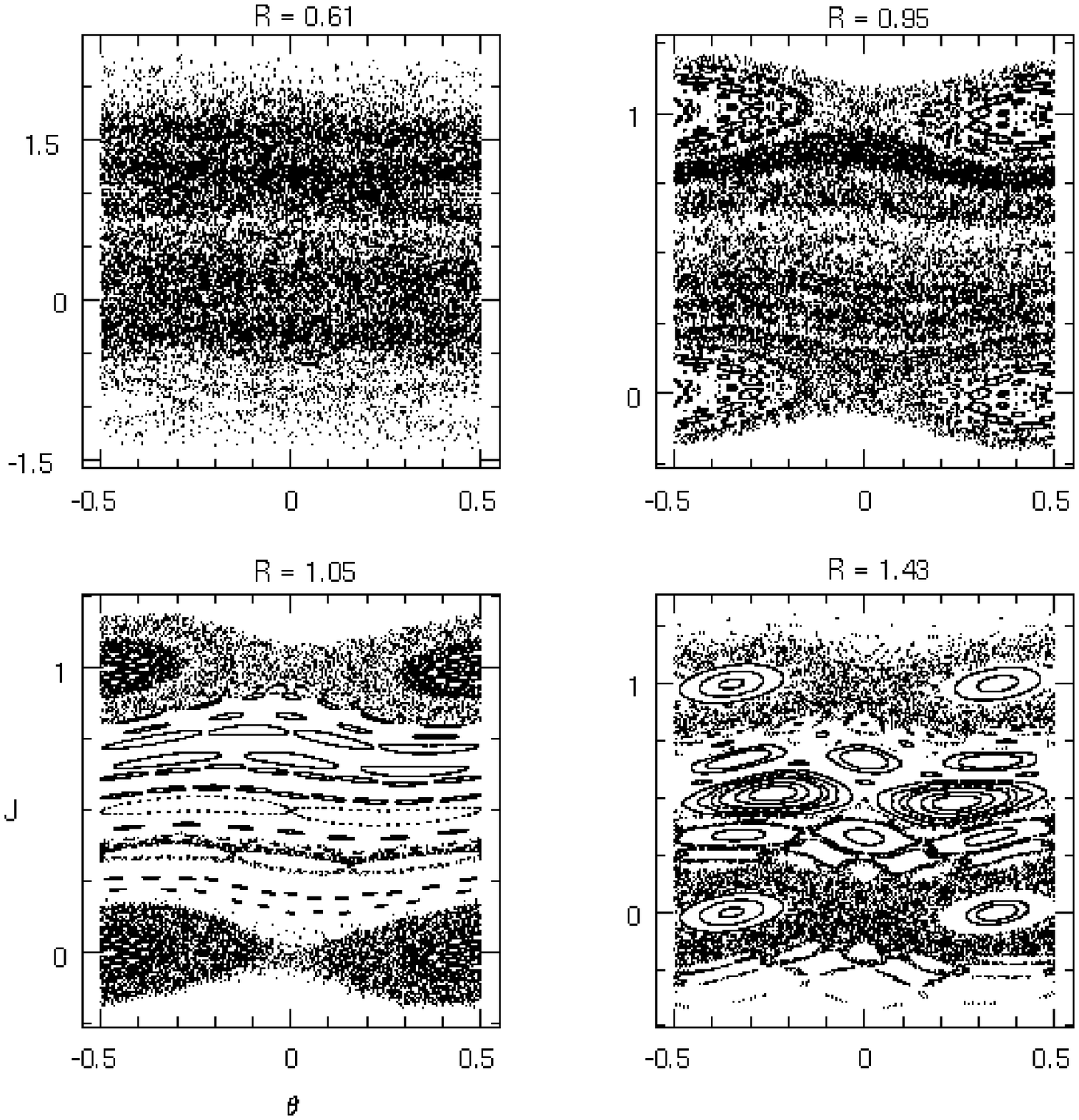}}
\caption{}
\label{xp}
\end{figure}

When $R$ assumes non-integer values the situation is entirely
different from the earlier case, particularly when $K$ is small. Shown
in Fig. \ref{xp} are typical phase space structures of the GSM for $R \neq
1$ and $0 < R < 2$ at small $K$. It is seen that no KAM tori exist
even when $R$ is slightly away from unity. In general, no KAM tori
appear in the phase space when $R$ departs from integer values.  This
is of course
due to the discontinuity in the map when $R$ assumes non-integer
values. In other words, there are no closed loops ${\cal C}$
encircling the cylinder such that ${\cal T'}{\cal C}={\cal C}$, where
${\cal T'}$ is the transformation defined by GSM with non-integer
values of $R$. As can be seen easily the discontinuity in ${\cal T'}$
arises at ${\theta}=-1/2~(\hbox{or}~1/2)$. 

This discontinuous dynamics
implies the abrupt breakup of KAM tori and a failure of the Poincar\'{e}-
Birkhoff scenario. The modifications that may occur are little
understood, and Chirikov in \cite{leshouches} points out that the piecewise
linear sawtooth map 
$(p_{n+1}= p_{n}+ K \, x_{n},\;\; x_{n+1}= x_{n}+p_{n+1} \;\mbox{mod}\;1)$
when $K$ is a non-integer such that $-4<K<0$ is one such system where 
extremely complicated locally stable motions occur and ``it is not at all
clear what could be a meaningful description, if any, of this apparently 
trivial model''. The particle in a well naturally gives rise to a non-linear
generalization of the sawtooth map and below we show how a simple
method may give us significant local stability information that explains
the dramatic differences observed in Fig. ~\ref{xp}.

In addition Fig. \ref{xp} distinguishes the dynamics between the cases
$R \la 1$ and $R \ga 1$. The former case is more chaotic and the phase space
is filled with unstable orbits, while many regular and chaotic regions
are seen in the latter. Thus the entire standard map may be regarded
as being poised at a point of bifurcation when regarded as a function
of the parameter $R$.  However, it is seen that when $K$ is large,
there is no such apparent difference in dynamics between the above two
cases since the phase space becomes chaotic and diffusive in both the
cases.

\subsection{Stability of the tori}

In the absence of the general KAM theorem for irrational tori and 
the Poincar\'{e}-Birkhoff theorem for rational tori, we expect to apply some
kind of local analysis to study deviations from integer values of
$R$. Indeed we will see that the integer values of $R$ could be
thought of as bifurcation points for entire orbits.  We noted earlier
that even for small values of $K$ no KAM tori exist in the phase space
of the GSM when $R$ departs from unity. To explain the stability
changes we look at the behaviour of a single irrational KAM torus,
which exists at $R=1$, as $R$ varies. A typical observation
(Fig. \ref{kam}) shows that KAM torus becomes unstable and chaotic for
$R \la 1$ while it is replaced by a chain of stable islands for $R \ga 1$.

We begin by performing a linear stability analysis for a KAM tori with
respect to $R$. We assume below that $K$ is small enough for such an orbit to 
exist. Let us denote a GSM (with $R=k$, positive integer) KAM orbit as 
$\{\tilde{\theta}_{n}, \tilde J_{n} \}$. This satisfies the following mapping:

\begin{equation}
\begin{array}{lll}
\tilde J_{n+1}&=&\tilde J_n+(K/2\pi)\sin(2\pi k \tilde{\theta}_{n}) \\[6pt]
\tilde{\theta}_{n+1}&=&\tilde{\theta}_{n}+\tilde J_{n+1}
\hspace*{0.6cm} (\mbox{mod}\;\;1).
\end{array}
\label{sm}
\end{equation}

\noindent 
Also consider an orbit $ \{\theta_n, J_n \}$ of the GSM with 
$R=k+\mu$ and $|\mu| \ll 1$, with the same initial conditions as the 
above KAM orbit. Let us introduce the differences
${\Delta}{\theta}_n={\theta}_n-\tilde{\theta}_n$; 
${\Delta}J_n=J_n-\tilde{J}_n$. Then it is possible to write an exact mapping
equation for $\Delta{\theta}$, $\Delta{J}$ as

\begin{equation}
\begin{array}{lll}
{\Delta}J_{n+1}&=&{\Delta}J_n+
(K/2\pi)\left\{\sin\left(2\pi (k+\mu) ({\tilde\theta}_{n} + 
\Delta\theta_{n})\right) - \sin(2 \pi k \tilde{\theta}_{n})\right\} \\[6pt]
{\Delta}\theta_{n+1}&=&{\Delta}\theta_n+{\Delta}J_{n+1}.
\end{array}
\label{del}
\end{equation}
Expanding in terms of ${\Delta}{\theta}_{n}$ and retaining first 
order terms leads to a time dependent force and the non-autonomous
linear set of equations:
\begin{equation}
\begin{array}{lll}
{\Delta}J_{n+1}&=&{\Delta}J_n +
K (k + \mu) \cos(2\pi (k + \mu) \tilde{\theta}_{n})\Delta{\theta}_{n} 
+ A_n \\[6pt]
{\Delta}\theta_{n+1}&=&{\Delta}\theta_n + {\Delta}J_{n+1}
\end{array}
\label{del1}
\end{equation} 
with
\[ A_n = {K \over 2\pi} \left\{\sin(2 \pi (k + \mu) \tilde{\theta}_{n}) -
\sin(2\pi k \tilde{\theta}_{n}) \right\}.\]

The behaviour of such linear non-autonomous equations
can be quite complex (compare for instance the Mathieu differential
equation which also arises in linear stability analysis).  For our
analysis on the stability of the tori we wish be intuitive and derive
rough but useful estimates.  Although we never expand $R$ about an
integer, here we assume that this excursion is small, so that we can
treat the non-autonomous stability equations perturbatively. Appealing
to the method of averaging we simply replace the time dependent force
by its time average. The limitation of this is pointed out further ahead. 

Also since the motion on the KAM torus is
ergodic in $\tilde{\theta}$, we replace the time average by an
equivalent space average with uniform measure.  
Thus 
\begin{equation}
\cos(2\pi R \tilde{\theta}_{n}) \approx g(R)\, =\, 
\int_{-1/2}^{1/2} \cos(2 \pi R x) \, dx
\label{av1}
\end{equation}
This procedure leads us to:

\begin{equation}
\begin{array}{lll}
{\Delta}J_{n+1}&=&{\Delta}J_n+(K/\pi)\sin(\pi (k + \mu) )\Delta{\theta}_{n} 
+ A_n \\[6pt]
{\Delta}\theta_{n+1}&=&{\Delta}\theta_n+{\Delta}J_{n+1}.
\end{array}
\label{lmap}
\end{equation}
 
\noindent Eqn.(\ref{lmap}) is a simple linear map and its stability can be 
seen from the corresponding Jacobian matrix
which gives the stability condition 
$|\;2+(K/\pi)\sin(\pi (k+\mu))\;|<2$.
This leads to the following window of stability: 
\begin{equation}
-{4\pi \over K} \;\; < \;\; {(-1)}^k \sin(\pi\mu ) \;\; < \;\; 0.
\label{win}
\end{equation}

The above stability window implies that if $k$ is odd KAM tori 
which exist at $R=k$ are stable for $\mu \ga 0$ 
and unstable for $\mu \la 0$; converse is the case if $k$ is even. 
Although $\mu$ is small in the above analysis, our 
numerical observations satisfy the stability condition (\ref{win}) even for 
large values of $\mu$ (see Fig. \ref{kam}, which is with $k=1$). This 
explains large phase space regions in Fig. \ref{xp}.

Before dwelling more in this region of phase space we divert attention
to those where this procedure has failed.  The chaotic regions around
$(J=0,\theta=0)$ in Fig.~\ref{xp} for instance presents such a
case. In the interior of the original $0/1$ resonance (at $R=1$) even
a reversal of the stability criteria is observed.  The hyperbolic
fixed point at $R=1$ persists for $R \ne 1$ and is the genesis of the
chaos. The smooth stable and unstable manifolds of the hyperbolic
fixed point that exist for $R=1$ and small $K$ cannot exist for $R \ne
1$ as they would then represent rotationally invariant
curves. Therefore the presence of the hyperbolic points necessarily
implies chaos. 

Within the framework of the stability analysis done earlier what has
failed around the hyperbolic fixed points is the assumption of uniform
measure. If we think of the simple pendulum we are moving from
energetic rotational motions (KAM) to slow oscillations of large amplitude and
clearly the time is spent preferentially around the turning points.
Also the replacement with time averages will fail as the time scales involving 
fast $\theta$ motions and slow $\Delta J$ motions become comparable.
The latter effect becomes important as we move into the $0/1$ resonance 
region. We note that the question of the existence of two different time 
scales is initial value dependent and is present in the standard map as well,
{\it i.e.,} it does not arise as a result of the presence of the additional 
parameter $R$.

The modified  measure can be approximated well by considering the simple 
pendulum. We now take $k=1$ for simplicity and consider only excursions
of $R$ from unity. The term
$\cos(2\pi R \tilde{\theta}_{n})$ in Eq. ~(\ref{del1}) is replaced 
with the ergodic average: 
\begin{equation}
\cos(2\pi R \tilde{\theta}_{n}) \approx g(R,\theta_c)\, =\, 
\int_{\theta_c}^{1/2} \frac{\cos(2 \pi R x) \, dx}{\sqrt{E\,-\, \cos(2 \pi x)}}
/ 
\int_{\theta_c}^{1/2}  \frac{dx}{\sqrt{E\,-\, \cos(2 \pi x)}}
\label{av2}
\end{equation}
Here $-1 \le E \le 1 $ is the scaled energy and $\theta_c$ is the turning 
point given by $\cos(2 \pi \theta_c)\,=\, E$.  
This is shown in Fig. ~\ref{avg2} and the point where $g$ crosses
zero from above ought to be a point where stability is recovered. 

This then explains the uniform chaos around the 
hyperbolic fixed point for $R\ne 1$. This also explains the 
stability around the wall $(\theta=1/2, J=0)$ in all cases.
However the interior of the resonance is not accessible to the
simple theory. Remarkably  Fig. \ref{avg2} predicts a recovery of stability
at $\theta_c \approx 0.15$ which is seen when $R=0.95$ while 
stability is recovered at a much higher value of $\theta_c$ for $R=1.05$ and
is not seen in  Fig. ~\ref{avg2}. This fact we attribute to the failure
of the averaging procedure in analyzing non-autonomous equations and
have no other recourse than Eq.~(\ref{del1}) itself.

As we have noted, hyperbolic fixed points generate chaos when $R\ne
1$.  The lack of chaos in the ``meat'' of the phase space for 
$R  \ga 1 $
where there were KAM tori are therefore regions where there is a total
absence of hyperbolic points. The Poincar\'{e}-Birkhoff theorem concerning
the breakup of rational tori into an equal number of 
hyperbolic and elliptic periodic points is clearly violated.  In fact
the modified scenario seems to be the creation of only elliptic fixed
points for $R \ga 1$ and only hyperbolic points for $R \la 1$. 
To illustrate this
we have shown in Fig. ~\ref{gm} the fate of points on the lines 
$J=1/3$ and $J=(\sqrt{5}-1)/2$. While in the former case three elliptic 
islands emerge prominently, in the latter one can see chains of elliptic
islands with the clear hierarchy of the number of islands $5,8,13,21,34 \dots$
deriving from the Fibonacci sequences generating closer approximations
to the golden mean. 

These island chains alternate as the ratios approach the golden mean.  
As a limiting case we may think of a golden mean chain, but we have 
of course not rigorously proved its existence.
There appear to be two types of orbits in the stable regions: those
that form an island chain and those that meander in the interstitial 
spaces between island chains on presumably a fractal set. These
would then be examples of strange non-chaotic sets in Hamiltonian 
mechanics. More investigations into these simple systems seems 
warranted.

We have thus acquired a rather detailed understanding, using 
simple methods,  of the 
dramatic stability changes that accompanies the introduction 
of an additional length scale or discontinuity. We may also
speculate that this may prove a good way of controlling
or enhancing chaos. While our study has been for Hamiltonian systems
dissipative systems may also display such a behaviour.  

\subsection {Lyapunov Exponents}

The Lyapunov exponents can be 
calculated using the Jacobian matrix of the GSM, which at $n$th iterate 
we denote as ${\bf M}_n$. The stability changes induced by $R$ 
ought to reflect on the exponent somewhat like a phase transition
effect on an order parameter.
Shown in Fig. \ref{le} is a typical calculation of the positive 
Lyapunov exponent ${\Lambda}_+$ for 
$K = 0.1\pi$. The orbit chosen is a KAM torus for $R=1$ and we, from our 
earlier discussion, expect it to be unstable for $R<1$ and stable for $R>1$ 
but not too large. It is seen that $\Lambda_+ \simeq 0$ for $R \ga 1$ in the 
range shown, which corresponds to the regular motion due to the stable 
islands formed from the orbit; for $R<1$, $\Lambda_+ > 0$ which indicates 
the chaotic behaviour of the orbit. Also the orbit has a maximum value 
of the Lyapunov exponent at $R \simeq 0.5$, and is hence maximally unstable at 
this point. We now explore this analytically.

The Jacobian has trace 
\begin{equation}
|\hbox{Tr}({\bf M}_n)| = |2 + KR \cos(2\pi R {\theta}_n)|
\end{equation}
\noindent 
and hence $|\hbox{Tr}({\bf M}_n)| > 2$ for $R<1/2$ as $|{\theta}_n| \le 1/2$. 
We remind the readers that this range of $\theta_n$ is the {\it entire} 
configuration space and $\theta_n \,=\, \pm 1/2$ represent the wall
boundaries. Since the Jacobian is only a function of the angle at a
given phase space point (and not also the action), we conclude that
the Jacobian has real eigenvalues {\it throughout} the phase space.
In other words, the system is hyperbolic for $R<1/2$.  This implies
that there are contracting and expanding real directions or
alternatively stable and unstable manifolds throughout phase space. It
is worth remarking that very few dynamical models are known exactly to
be completely chaotic or hyperbolic. These include the sawtooth maps
\cite{percival}, the baker maps, flows on surfaces of constant
negative curvature. The standard map even for large values of the
parameter $K$ is not proven to be completely hyperbolic.  Thus in this
context, we place the fact that GSM has a parameter range for which it
is completely hyperbolic for all positive values of the parameter $K$.

The similarity of the GSM, when $R<1/2$, with the piecewise linear sawtooth 
maps is made clear by the following linear approximation. We approximate the 
force $f({\theta}_n) = \sin(2 \pi R {\theta}_n)$ in the GSM by a monotonic
linear function $2\sin(\pi R){\theta}_n$. 
This leads to $|\hbox{Tr}({\bf M}_n)| \approx |2 + (K/\pi)\sin(\pi R)|$. 
The Lyapunov exponents in this approximation are given by
\begin{equation}
{\Lambda_\pm} = \ln (r \pm \sqrt{r^2 -1}) 
\label{le_approx}
\end{equation}
\noindent where $r = 1 + (K/2 \pi)\sin(\pi R)$.
It is clear from Fig. \ref{le} that the approximated exponent 
fits fairly well. Similar agreement was seen for a wide range of initial
conditions and different values of $K$
as well. This shows the validity of the above linear
approximation for $f({\theta}_n)$ when $R<1/2$, which enables us to
understand the gross behaviour of the Lyapunov exponents in this
regime. We note that the usual linear approximation 
$f(\theta_n) \approx 2\pi R\theta_n$ 
near the origin was found not to be as good as the above approximation.

Although this study has been essentially that of another area-preserving map,
we feel that these novel effects are worth emphasizing, especially since the
standard map has been realized in the laboratory in the form of kicked cold
sodium atoms, and the transitions to chaos in the quantum regime have been
studied as noted above. Implementation of the competing length scales in this 
experiment could lead to novel observable transport effects. Theoretical 
studies of the quantum dynamics of this system are currently underway.

\newpage 
{\large Figure Captions}\\

\begin{figure}[ht]
\centerline{\psfig{figure=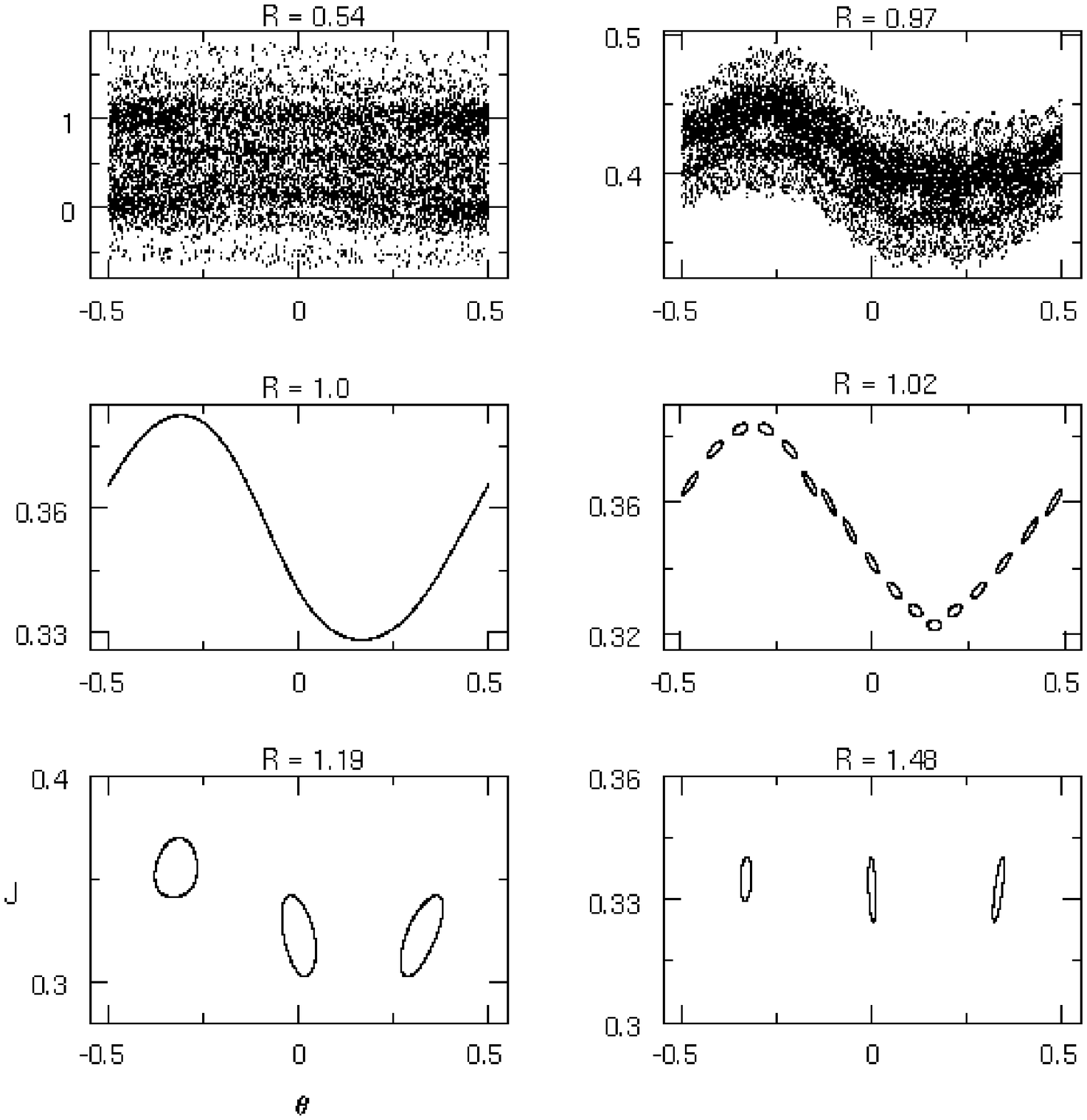}}
\caption{}
\label{kam}
\end{figure}

\begin{figure}[ht]
\centerline{\psfig{figure=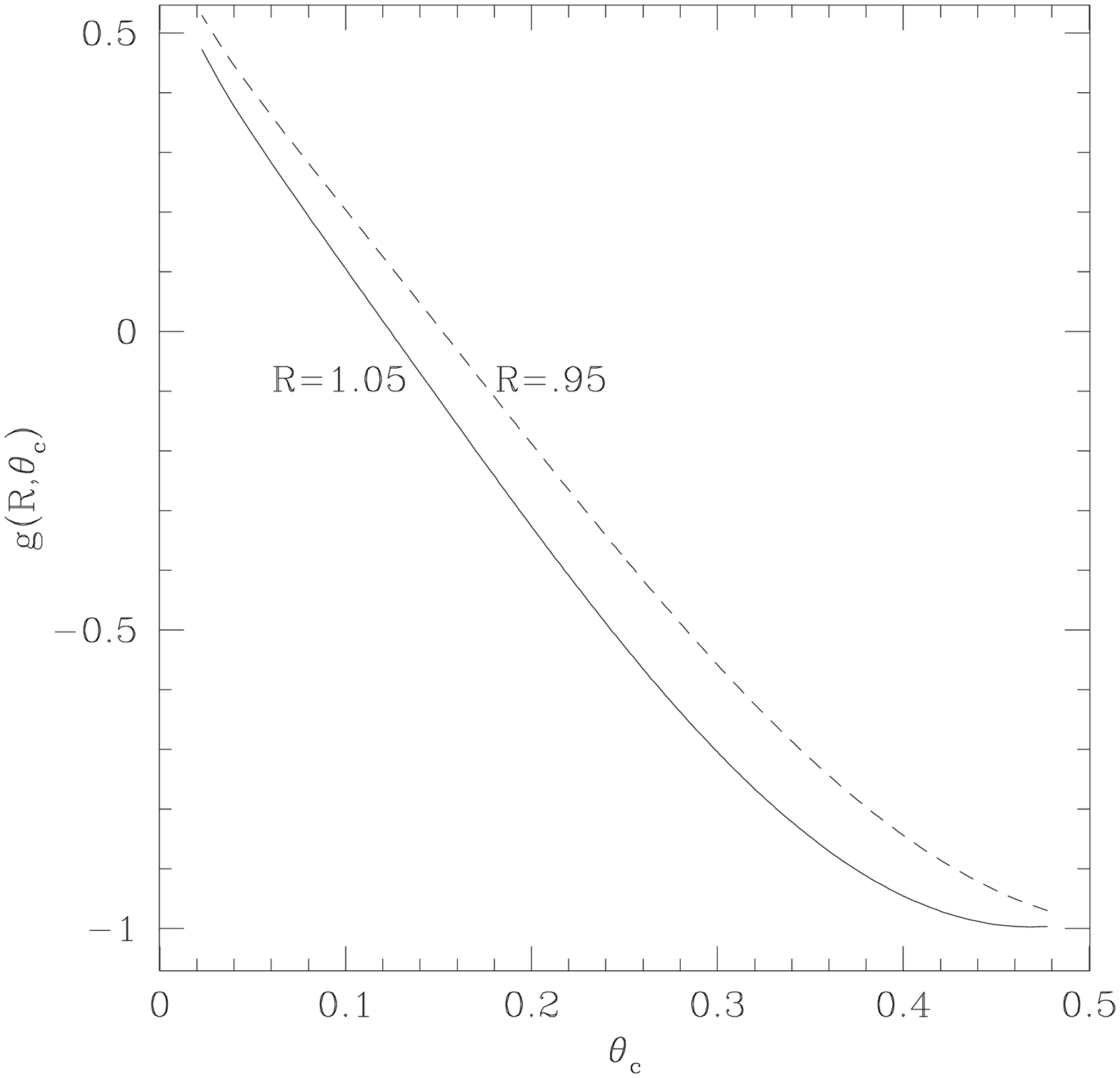}}
\caption{}
\label{avg2}
\end{figure}

\begin{figure}[ht]
\centerline{\psfig{figure=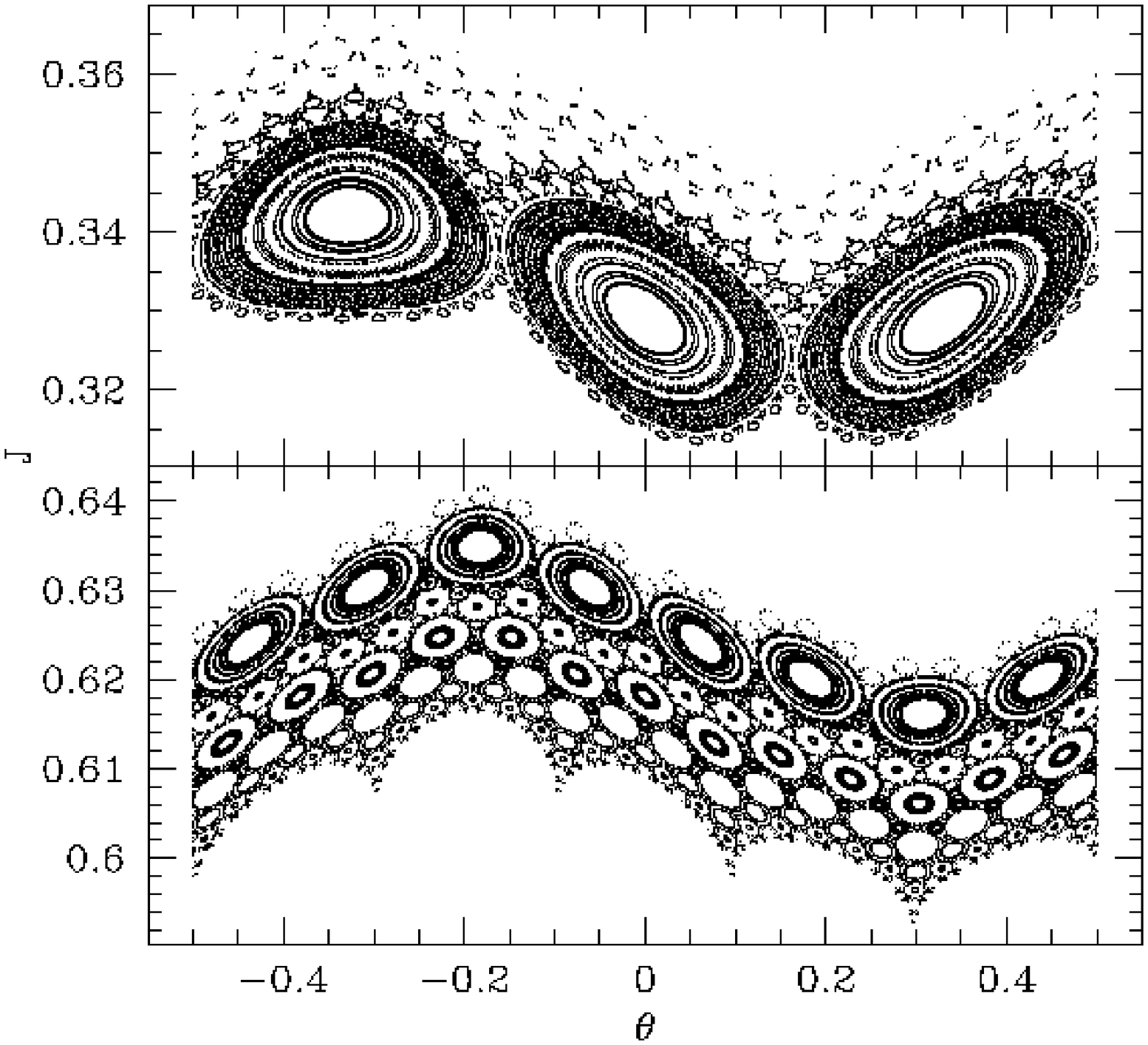}}
\caption{}
\label{gm}
\end{figure}

\begin{figure}[ht]
\centerline{\psfig{figure=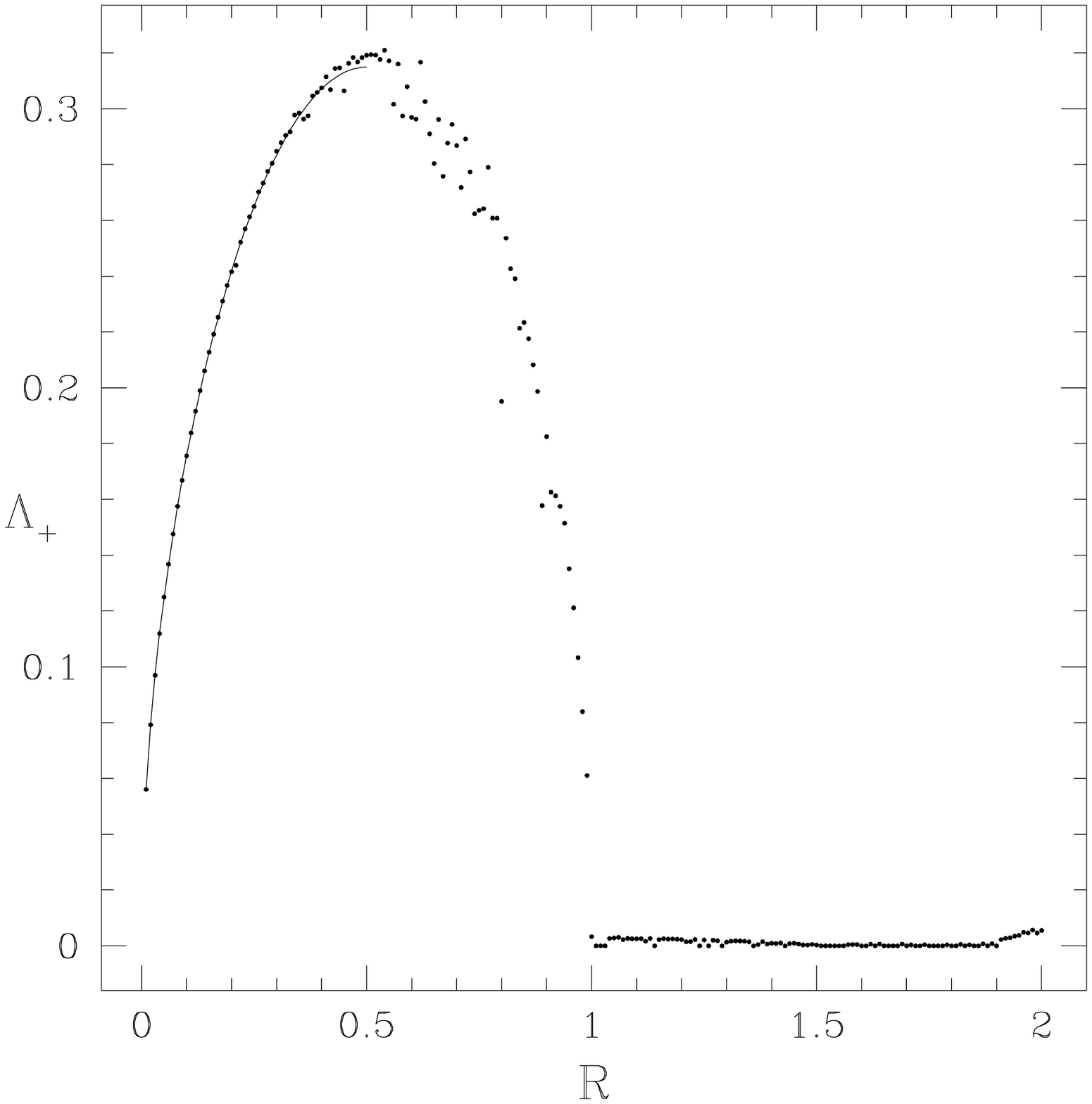}}
\caption{}
\label{le}
\end{figure}

\noindent
Fig 1 : Phase space of the GSM (with $R\neq 1$) at $K=0.1\pi$.
No KAM tori are seen in any of the cases.\\
Fig 2 : Shown is an orbit with initial condition $S_0=(0,0.34)$,
at $K = 0.1\pi$, which corresponds to a KAM torus when $R=1$. For $R \la 1$,
the orbit becomes unstable and chaotic, while for $R \ga 1$ it becomes
chain of stable islands.\\
Fig 3 : The function $g(R,\theta_c)$ which when positive
determines the instability of some GSM orbits.\\
Fig 4 : Orbits with initial conditions on the line
$J=1/3$ (top) and $J=(\sqrt{5}-1)/2$ (bottom) for $K=0.1$ and $R=1.05$.\\
Fig 5 : Positive Lyapunov exponent (dots) of the orbit shown in
Fig. \ref{kam}. The exponent is positive for $R<1$ with maximum at
$R \simeq 0.5$ and zero for $1<R<2$. Solid curve is the
approximation (\ref{le_approx}) for $R<1/2$.

\end{document}